\documentclass[pra,aps,twocolumn,tightenlines,showpacs]{revtex4}
\usepackage{epsfig}
\begin{document}

\title{Observation of light dragging in rubidium vapor cell}

\author{Dmitry Strekalov}
\author{Andrey B. Matsko}
\author{Nan Yu}
\author{Lute Maleki}

\affiliation{Jet Propulsion Laboratory, California Institute of
Technology, 4800 Oak Grove Drive, Pasadena, California 91109-8099}
\date{\today}

\begin{abstract}
We report on the experimental demonstration of light
dragging effect due to atomic motion in a rubidium vapor
cell. We found that the minimum group velocity is achieved for light
red-shifted from the center of the atomic resonance, and that the value
of this shift increases with decreasing group velocity, in agreement with
the theoretical predictions by Kocharovskaya,
Rostovtsev, and Scully [Phys. Rev. Lett. {\bf 86}, 628 (2001)].
\end{abstract}
\pacs{42.50.Gy, 03.30.+p}

\maketitle

The phenomenon of light dragging by a moving medium has a two century long
history \cite{paulibook,LLbook}. Starting with Fresnel's ether
theory \cite{fresnel18acp} and Fizeau's observation of light
dragging \cite{fizeau51cras}, several investigations of this effect 
were performed before the advent of relativity.
Later, the special theory of relativity greatly advanced the
understanding of this effect through
the relativistic addition of velocities
\cite{paulibook}. At the same time when the theory of relativity was
being developed, Lorentz \cite{lorentz95book} predicted the influence of
dispersion in a moving medium on the speed of light. This
effect was first verified by Zeeman who used glass rods for high 
dispersion placed on a moving platform \cite{zeeman14raa}.

The experimental approach to the study of light dragging by a moving 
medium has recently received a new opportunity, since it was found 
that coherent media can
have a huge dispersion that can result in ultraslow group velocities
of light (for review see
\cite{matsko01aamop,boyd02po}). The value of this dispersion is
many orders of magnitude greater than the dispersion of
glass used by Zeeman, for example. Ultraslow group velocities have been
observed in atomic vapors \cite{hau99nature,kash99prl,budker99prl}
as well as in doped solids \cite{turukhin02prl,bigelow03prl}. These
results have revealed new and exciting opportunities for the study of
fundamental phenomena involving light dragging, for instance,
by suggesting dielectric analogies of astronomical effects
\cite{leonhardt00prl}.

The frequency dispersion responsible for the ultraslow light
propagation occurs in the vicinity of a narrow transparency window
resulting from Electromagnetically Induced
Transparency (EIT) \cite{arimondo96,harris97pt,marangos98jmo}.
The specific structure of the EIT medium has raised questions about the
minimum group velocity of light that can be realized. The 
ponderomotive effects \cite{artony01jetpl} as well as light
reflection from the medium boundary and absorption in the
dielectric have been identified as possible sources of influence on 
the achievable minimum group velocity \cite{kozlov02pla}.

It was also shown that it is possible to slow down
a light pulse or even bring it to a full stop using the effect of
light dragging \cite{kocharovskaya01prl}. The light pulse can be at rest
with respect to an observer if the medium moves uniformly with a
velocity equal and opposite to the group velocity of
light propagating in the stationary medium. Moreover, light dragging was
predicted \cite{kocharovskaya01prl} in an atomic cell
containing hot atomic vapor. In this case the effect was predicted to 
be achieved by selecting a proper velocity
subgroup in the inhomogeneously (Doppler) broadened medium. The
selection can be realized by tuning the frequency of a drive
laser and is nothing more than spectral hole-burning,
a phenomenon well known in nonlinear spectroscopy \cite{letokhovbook}.

The experimental realization of ``light freezing" and
the observation of light dragging in a Doppler-broadened medium
nevertheless encountered difficulties \cite{mikhailov03jmo}. This is because
the use of either optically thick collimated atomic
beams, or vacuum atomic cells is required. The small group velocities of
light, on the other hand, was realized in hot atomic cells
containing a buffer gas, or in coated atomic cells. The presence of a 
buffer gas
or coating results in velocity mixing of the atoms via
velocity changing collisions. It becomes impossible to select a
proper velocity subgroup as discussed in \cite{kocharovskaya01prl}.

We report on an experiment for the observation of light dragging
in a hot vacuum atomic rubidium cell without any butter gas or coherence
preserving coating. We realized the experiment by a proper tuning of
the experimental parameters which allowed us to reach $V_g \approx
1.7$~km/s average group velocity in a $L=5$~cm long cell. We found
that the group velocity decreases with red tuning of the drive
laser radiation (which corresponds to selecting of an effective ``atomic beam''
moving in the opposite direction with respect to the light),
and increases with blue tuning. As a result, the maximum of
average group delay for light propagating in the cell shifts
as much as $40$~MHz to the red. The following theoretical analysis
expands on our results.

An understanding of light dragging can be gained from Lorentz
transformation of the wave vector and frequency of electromagnetic
waves. Using this transformation and the definition of group velocity, we
come to the conclusion \cite{LLbook} that if light propagates in
the direction of the medium flow, its group velocity in
laboratory frame, $\widetilde V_g$, changes as
\begin{equation} \label{dragging}
\widetilde V_g = V_g+ v \left(1-\frac{V_g^2}{c^2} \right
)-\frac{vn\omega_0}{c} \frac{dV_g}{d\omega_0},
\end{equation}
where $V_g$ is the group velocity in the frame of reference moving
with the medium, $v$ is the velocity of
the medium relative to the laboratory frame of reference, $c$ is
the speed of light in the vacuum, $n$ is the index of refraction
of the medium, $\omega_0$ is the frequency of the
electromagnetic wave.

Eq.(\ref{dragging}) is general; it does not depend on the specific 
properties of a medium. Hence, it is valid for a hot atomic vapor 
where small
group velocities could be realized. The goal of the present paper
is to detect light dragging based on Doppler effect in an atomic
cell.

The velocity of an atomic subgroup (that can be selected by
standard techniques of nonlinear spectroscopy \cite{letokhovbook})
should be compared to the average thermal
speed $ v_T  = (2kT/M)^{1/2}$ of atoms in the cell, where $k$ is
Boltzmann constant, $T$ is the temperature, and $M$ is the atomic
mass. For $^{87}$Rb, this speed is equal to approximately
$230$~m$/$s for $T=65{\rm ^oC}$. The
minimum group velocity observed in our experiment is much larger
than this value. To detect the light dragging effect we measure the shift
of the minimum group delay with respect to the laser tuning that
determines the atomic velocity subgroup.

\begin{figure}[ht]
  \center{
\epsfig{file=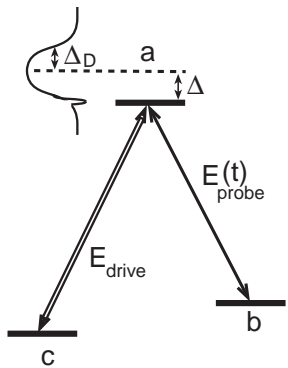, width=3.8cm, angle=0} }
  \caption{\label{fig:lambda}Simplified energy levels scheme.
The drive and probe fields interact with a velocity subgroup of atoms
moving at $v_T\Delta/\Delta_D$. The velocity width of this effective
``atomic beam" is
$\Delta v\approx\frac{2|\Omega_{drive}|}{k_{drive}}
\sqrt{\frac{\gamma_{ac}}{\gamma_{cb}}}
\ll v_T$ \cite{kocharovskaya01prl}, where $|\Omega_{drive}|$
is the Rabi frequency of the drive, $\gamma_{ac}$ is the homogeneous
decay rate of the drive transition, $\gamma_{cb}$ is the coherence
decay rate, $k_{drive}$ is the drive wave vector.}
  \end{figure}

Let us estimate this shift. We consider a three level $\Lambda$
configuration shown in Fig.\ref{fig:lambda}. Such a scheme is very
convenient for understanding of EIT based on coherence population
trapping \cite{arimondo96}, though it is
not applicable for quantitative description of the effect in real
atoms \cite{greentree03}. Drive and probe
lights interact with transitions $|a \rangle \rightarrow
|c\rangle$ and $|a \rangle \rightarrow |b\rangle$, respectively.
The dipole-forbidden transition $|c \rangle \rightarrow |b\rangle$ has
a long life time $\gamma_{bc}^{-1}$ determined primary by the
interaction time of atoms and light.

%%%%%%%%%%%%%%%%%%%%%%%%%%%%%
%%%%%%%%%%%%%%%%%%%%%%%%%%%%%

The dependence of the group velocity on the drive frequency in the 
motionless medium may be estimated from
\begin{equation}
V_g = V_g(0) \left [1+ \beta(T)\left (\frac{\Delta}{\Delta_D}
\right )^2 \right ],
\end{equation}
where $\Delta = \omega_0 - \omega_{ac}$ (Fig.(\ref{fig:lambda})),
$\Delta_D = 215$~MHz is the half width at half maximum of the
Doppler profile, $\beta(T)$ is a function of temperature that
is used to take into account propagation and optical pumping
effects. For optically thin medium consisting of $\Lambda$ atoms
$\beta(T) = 1$ \cite{kocharovskaya01prl}. 
This function can be determined by comparing measured distribution 
widths to $\Delta_D$.

The drive laser detuned from one-photon resonance by $\Delta$ makes
the probe interact with the atomic ``beam".  This ``beam" has an 
average speed $v=v_T
\Delta/ \Delta_D$ while moving in the direction of the probe wave.
According to (\ref{dragging}) the group velocity with respect to
the laboratory frame of reference is approximately equal to
\begin{equation}
\widetilde V_g \approx V_g(0) \left [1+ \beta(T)\left
(\frac{\Delta}{\Delta_D} \right )^2 \right ]+ v_T
\frac{\Delta}{\Delta_D}.
\end{equation}
The group velocity reaches a minimum at detuning $\Delta_{opt}$
\begin{equation}\label{DeltaRel}
\frac{\Delta_{opt}}{\Delta_D} = -\frac{ v_T }{2 \beta(T)V_g(0)}.
\end{equation}
This minimum corresponds to the maximum of group delay
$L/\widetilde V_g$. The frequency shift of the maximum group
delay could be detected and measured.

\begin{figure}[ht]
  \center{
\epsfig{file=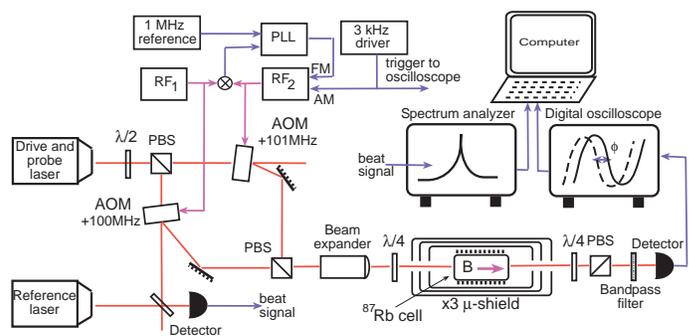, width=9.5cm, angle=0} }
  \caption{\label{fig:setup}A simplified diagram of the experimental 
setup. RF$_1$ and  RF$_2$ are
the drivers for the acousto-optical modulators (AOM) that are locked
to an external 1 MHz reference by a phase-lock loop (PLL);  PBS are the
polarizing beam splitters. }
  \end{figure}

The basic elements of our experimental setup include a rubidium
atomic vapor cell with an enriched amount of $^{87}$Rb isotope;
a frequency stabilized diode laser producing the drive and probe beams;
and a reference laser stabilized with a reference rubidium cell
allowing us to measure the exact frequency of the radiation passing
through the atomic cell by observing a beat signal. The amplitude of 
the probe laser
is modulated at $3$~kHz with an acousto-optical modulator.
We measure the delay of the beat-note of the probe radiation to detect
the light dragging phenomenon. Let us now describe the setup in
more detail (see Fig.\ref{fig:setup}).

%%%%%%%%%%%%%%%%%%%%%%%%%%%%%
%%%%%%%%%%%%%%%%%%%%%%%%%%%%

The two-inch-long atomic cell contains 75\% of $^{87}$Rb and 25\%
$^{85}$Rb. The cell can be kept at a constant temperature ranging
from $45^o$C to $75^o$C, stabilized to $0.1^o$C with a feedback loop.
The cell is placed inside
a solenoid, which is used to create a bias magnetic field along the
propagation direction of the optical beam. This field introduces a 
$1$~MHz Zeeman detuning
for the ground state sublevels with $\Delta m = 2$. The
solenoid and the cell are enclosed in a three-layer $\mu$-metal
shield to cancel stray magnetic fields.

We use a Vortex laser system tuned to the $5{\rm S}_{1/2}$, F=2
$\rightarrow\; 5{\rm P}_{1/2}$, F=1 transition of $^{87}$Rb. The
laser radiation is split into two parts: drive and probe,
whose powers can be varied, and their beam radius is set at
$R=5$~mm by the beam expander.

The short term stability of our laser is better than
$1$~MHz/Hz$^{1/2}$. However, the long term stability is rather
unsatisfactory, so for the accurate frequency measurements we had to use a reference
laser locked at the crossover
resonance at $^{85}$Rb D1 line in an additional cell
(not shown in Fig.\ref{fig:setup}). Measuring the beat frequency of these two
lasers with a spectrum analyzer allowed us to track the detuning of the
first laser from the transition with better than $3$~MHz resolution
during up to five hours, when the first laser was stepped through
$2$~GHz range several times.

To split the laser beam into the drive and probe we use a polarizing 
beam splitter (PBS)
preceded by a half-wave plate; then, each beam passes through an 
acousto-optical modulator
(AOM) up-converting one frequency by $100$~MHz, and the other by $101$~MHz.
The $1$~MHz frequency difference between the two beams is enforced
by phase-locking the AOM drivers' beat signal to a $1$~MHz reference
signal in the master-slave phase lock loop (PLL) configuration, and
corresponds to $\omega_{bc}$ in Fig.~(\ref{fig:lambda}).
The AOM in the probe (the slave)
channel also performs the function of modulation of the probe 
amplitude at $3$~kHz.
The phase of this modulation is later used to determine the group 
delay of the probe light.

As a result, we obtain the CW drive and amplitude modulated probe
beams with orthogonal linear polarizations, and with the frequency
detuning exactly matching the ground state splitting for $\Delta m = 2$
imposed by the solenoid field. To couple these ground states, the 
linearly polarized  drive
and probe fields are transformed to opposite circular polarizations 
by a quarter wave plate.
This step occurs after they have been combined by the second PBS and 
passed through the beam expander
to maintain a good quality Gaussian beam of the prescribed radius.

Light that passes through the cell is converted back to linear basis with
the second quarter wave plate, and then the drive light is rejected by
a polarizer. The extinction ratio is approximately 0.5\%. Probe 
radiation is detected by a
photodiode whose output is coupled into a digital oscilloscope,
to monitor the signal. We use the phase measurement function of the 
oscilloscope to
determine the group delay of the $3$~kHz beat-note, and the amplitude 
measurement
function to measure its transmission. Note that this type of 
measurement completely
rejects the drive leaking into the probe polarization (e.g. due to 
polarization rotation)
and provides us with very clean probe absorption data.

Let us turn now to the measurement results. The minimum full width
at half maximum of the EIT resonance that we detected is $20$~KHz,
which is slightly narrower than the inverse
average transit time of atoms through the laser beam $v_T/(2R)$,
for the beam diameter used in our setup. The additional narrowing
of the resonance could orginate from the density narrowing effect of
the EIT resonance \cite{lukin97prl}.

The maximum group delay of the $3$~KHz beat note was approximately
$30~\mu$s. This group delay was detected for a $72^o$C cell
temperature, $2.4$~mW drive and $0.15$~mW probe field powers.
Drive and probe fields had approximately 30\% and 10\% relative transmission
respectively.

\begin{figure}[ht]
  \center{
\epsfig{file=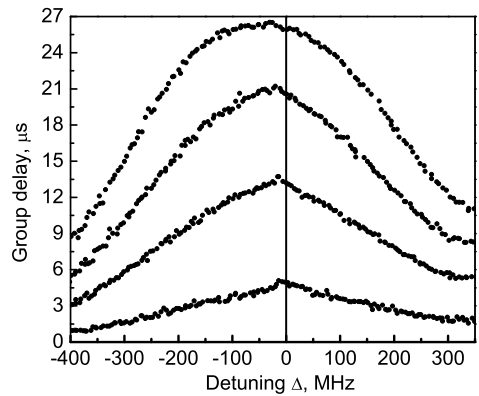, width=7.cm, angle=0} }
  \caption{\label{fig:DelaySampl}Typical dependence of the group delay on
the single photon detuning $\Delta$. The curves correspond to (top to bottom)
$T= 72,\,68,\,62$ and $49{\rm ^oC}$. }
  \end{figure}

We first detected the group delay for the beat-note of the probe at a
couple of dozens of cell temperatures. The data were taken with
a fixed two-photon detuning, drive and probe powers, and varying
one-photon detuning. Four typical data curves are shown in
Fig.\ref{fig:DelaySampl} (maximum delay increases with temperature). It
is easy to see that the curve shapes are non-Lorentzian for any
temperature. Moreover, the width of the dependencies increases
almost twice as fast with the temperature increase from 45$^o$C to
72$^o$C. Such a shape is determined by the multilevel structure of
the system as well by the fact that we study an open transition.

%%%%%%%%%%%%%%%%%%%%%%%%%%%%

\begin{figure}[ht]
  \center{
\epsfig{file=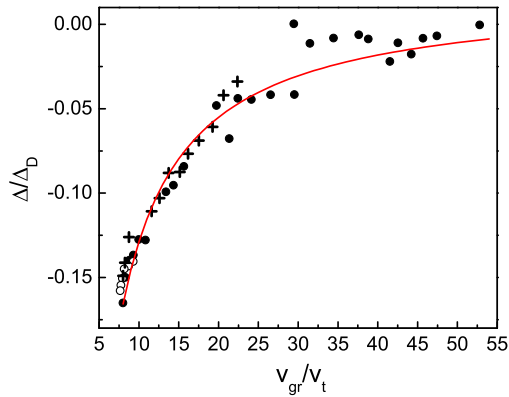, width=7.5cm, angle=0} }
  \caption{\label{fig:ShiftvsV}Combined results showing the 
dependency of the detuning
$\Delta$ corresponding to minimum group velocity $v_{gr}$ on $v_{gr}$.
The data are obtained by: fixing the drive power at 2.4 mW (solid 
dots) and 2.8 mW (crosses)
and varying temperature from 45$^o$C to
72$^o$C; and by fixing temperature at 69$^o$C  and varying the drive 
power (empty circles).
The solid line is the theory curve obtained from Eq.(\ref{DeltaRel}).}
  \end{figure}

It is easy to see that the center of curves in
Fig.\ref{fig:DelaySampl} are shifted each with respect to the other. At low
temperatures the center coincides with the resonant frequency of
atomic transition, while for large temperatures it shifts to the
red. We found the maximum delay and shift of the point of maximum delay for
each curve. To ensure consistency of the results we repeated the experiment
with a different drive power (2.8 mW). The combined measurement 
results are shown
in Fig.\ref{fig:ShiftvsV} by dots and crosses for different drive powers. We
also performed measurements of the group delay versus power of the
drive radiation at a fixed temperature $T = 69{\rm ^oC}$ and placed 
the data on the
same plot as empty circles. The dependencies perfectly agree with
each other, confirming the theoretical prediction of 
Eq.(\ref{DeltaRel}), i.e. the
frequency shift of the maximum group delay only depends
on the maximum delay itself and not on how it is achieved.

%%%%%%%%%%%%%%%%%%%%%%%%%%%%

To compare the theoretical predictions and experimental data we plot 
the function
given by Eq.(\ref{DeltaRel}) as a solid line in 
Fig.\ref{fig:ShiftvsV}. In this function,
the width parameter $\beta(T)$  is determined from the widths of experimental
curves such as shown in Fig.\ref{fig:DelaySampl} and varies from unity for
the low temperature curves to approximately one half for the high temperature curves.

In conclusion, we have demonstrated the effect of
light dragging occurring in a hot atomic vapor due to thermal
motion of atoms. Our experiment confirms the influence of
spatial dispersion of the refractive index of a medium on light
propagation. 
Moreover, we have measured the group velocity dragging for a beat-note 
of  monochromatic waves, which is equivalent to a  measurement of differential
phase velocity dragging, and is an indirect verification of 
Lorentz theory for phase velocity of light in moving media.
Our results also show that coherent atomic media really
possess large light dragging abilities so the ideas to
``freeze light" with such a motion are reasonable.

The research described in this paper was carried out under
sponsorship of DARPA by the Jet Propulsion Laboratory, California
Institute of Technology, under a contract with the National
Aeronautics and Space Administration.

%%%%%%%%%%%%%%%%%%%%%%%%%%%%%%%%%%%%%%%%%%
%%%%%%%%%%%%%%%%%%%%%

\end{document}